# Spin resonance of 2D electrons in a large-area silicon MOSFET

S. Shankar, A. M. Tyryshkin, S. Avasthi, S. A. Lyon

Department of Electrical Engineering, Princeton University, Princeton, New Jersey 08544, USA

Corresponding author: S. Shankar. Tel: +1-609-258-6157; E-mail: sshankar@princeton.edu

**Abstract:**

We report electron spin resonance (ESR) measurements on a large-area silicon MOSFET. An ESR signal at g-factor 1.9999(1), and with a linewidth of 0.6 G, is observed and found to arise from two-dimensional (2D) electrons at the $Si/SiO_2$ interface. The signal and its intensity show a pronounced dependence on applied gate voltage. At gate voltages below the threshold of the MOSFET, the signal is from weakly confined, isolated electrons as evidenced by the Curie-like temperature dependence of its intensity. The situation above threshold appears more complicated. These large-area MOSFETs provide the capability to controllably tune from insulating to conducting regimes by adjusting the gate voltage while monitoring the state of the 2D electron spins spectroscopically.

Quantum computing and spintronics proposals have suggested utilizing electron spins near semiconductor interfaces, in the hope that electrons can be moved freely and controllably (using electrical gates) near the interface while preserving their spin states [1]. Electron spin resonance (ESR) is the spectroscopic tool of choice to probe spin states of electrons, however severe sensitivity requirements (at least $10^9$ spins) makes it difficult to use ESR to measure spins at semiconductor interfaces [2]. Recently, ESR has been successfully demonstrated on 2D electrons in Si/SiGe [3] and Si/SiC [4] heterostructures, however previous measurements of 2D electrons at the $Si/SiO_2$ interface were only below threshold [5]. Here we report ESR measurements of 2D electrons in MOSFETs both above and below threshold. Our MOSFET devices provide the ability (by adjusting the gate voltage) to arbitrarily set the electron density at the interface and thus to study spin states both in the insulating and conducting regimes using the same device.

Large area MOSFETs (gate area of $0.4 \times 2$ cm$^2$) were fabricated to obtain adequate ESR signal from 2D electrons. The device area of about 1 cm$^2$ (assuming an electron density $\sim 10^{11}$ cm$^{-2}$, and 1% spin polarization at X-band ESR magnetic fields) provides at least $10^9$ unpaired electron spins, above the detection limit of X-band ESR experiments. Inversion MOSFETs were fabricated on a Si(100) wafer (boron doped at $10^{15}$ cm$^{-3}$), using standard techniques. The devices consisted of phosphorus implanted source-drain contacts, $\sim$100 nm of dry thermal oxide and a Ti/Au metal gate. Accumulation MOSFETs were fabricated using identical steps on a Si(100) 7 μm epi-wafer (phosphorus doped at $10^{15}$ cm$^{-3}$). A typical MOSFET gave a threshold voltage of 1.1V at 4.2 K measured from *I-V* curves.

ESR measurements were performed at temperatures between 4 and 20 K using an X-band ESR spectrometer (Bruker Elexsys580). For the electrically detected magnetic resonance (EDMR) experiment, the spectrometer setup was modified to measure changes in resistance of the MOSFET while exciting ESR transitions [6, 7].

ESR spectra for the inversion MOSFET, measured at 5 K and at three gate voltages ($V_G$), are shown in Fig. 1. A weak signal at $g = 1.9988(1)$ arises from the implanted source-drain areas in the MOSFET as evidenced by an increase of the signal intensity when the source-drain areas were positioned closer to the center of the resonator. This signal, which shows no dependence on $V_G$, arises from conduction electrons in the degenerately doped source-drain contacts [8, 9].

In contrast, as seen in Fig. 1, a stronger signal at $g = 1.9999(1)$ shows a pronounced dependence of its intensity on $V_G$. This gate voltage dependence identifies the signal as arising from electrons located in the gated area of the MOSFET. An identical signal was observed in the accumulation MOSFET. The g-factor, 1.9999, is close to that reported for 2D conduction electrons in Si/SiGe heterostructures [7, 10] and also for 3D electrons in bulk silicon [9]. Further, an EDMR experiment above threshold (bottom trace in Fig. 1) reveals an identical signal at $g = 1.9999(1)$. Therefore we assign the signal above threshold to 2D conduction electrons in the MOSFET.

When $V_G$ is reduced below threshold (e.g., $V_G = 0$ V in Fig. 1) the signal shows no visible change in linewidth and a small decrease in g-factor ($\sim 1 \cdot 10^{-4}$). These small changes in g-factor and linewidth suggest that electrons measured below threshold, though immobile, are similar in character to mobile conduction electrons. For example, these electrons might be weakly confined by potential fluctuations or shallow traps at the $Si/SiO_2$ interface [11].

The number of unpaired spins ($N_S$) in the MOSFET, calculated by double integrating the derivative ESR signal at $g = 1.9999$, is plotted as a function of $V_G$ in Fig. 2. $N_S$ increases monotonically with an increase in $V_G$, and saturates at high electron densities ($V_G > V_{Th}$). Similar behaviour was reported for 2D electrons in gated Si/SiGe quantum wells [3], with a saturation in the spin density ($N_S$ per unit area) observed at high electron densities. In the limit of low temperatures and high electron densities ($k_BT \ll$ Fermi energy, $E_F$), $N_S$ for an ideal, non-interacting 2D electron gas (2DEG) is proportional to the density of states (DOS) at the Fermi energy. The DOS of a 2DEG is independent of $E_F$, hence $N_S$ is also independent of $E_F$ at high electron densities. Thus, the saturation of $N_S$ for $V_G$ above threshold in Fig. 2 supports our conclusion that the signal at $g = 1.9999$ is from 2D conduction electrons in the MOSFET.

For $V_G$ below threshold, $N_S$ measures the number of unpaired spins arising from weakly confined electrons as suggested above. As $V_G$ is reduced below threshold, $N_S$ at first decreases, and then (below 0.4 V at 5 K) remains constant. At these lowest gate voltages, $E_F$ is well below the conduction band edge ($E_C$ - $E_F$ > $k_BT$), and therefore trapped electrons are unable to thermally escape to the conduction band; electrons are trapped in quasi-equilibrium states, making $N_S$ independent of $V_G$. However, as seen in Fig. 2 (▼), illuminating the sample at low $V_G$ helps the system reach equilibrium, by generating holes that neutralize charges at the Si/SiO$_2$ interface, with the net result that $N_S$ falls significantly.

The temperature dependence of $N_S$ above and below threshold, shown in Fig. 3, further clarifies the origin of the signal at $g$ = 1.9999. At $V_G$ = 0.8 V, $N_S$ scales as $1/T$, thus revealing a Curie-like susceptibility dependence. Since the Curie law is characteristic of isolated, independent electrons, the data supports our assignment of the ESR signal below threshold to confined electrons at the MOSFET interface.

In contrast, $N_S$ measured above threshold ($V_G$ = 2 and 3 V in Fig. 3) shows a different behaviour. 2D conduction electrons are expected to show Pauli susceptibility. The dashed and dotted curves in Fig. 3 show the Pauli dependences calculated for the electron densities of $3.9\times10^{11}$ cm$^{-2}$ ($V_G$ = 2 V) and $6.3\times10^{11}$ cm$^{-2}$ ($V_G$ =3 V), measured for this MOSFET in a Hall experiment at 4.2 K. The Pauli dependences make a poor fit to the experimental data especially at 3 V and at higher temperatures. Apparently the Pauli dependence alone can not fully explain the data.

In conclusion, we have fabricated and performed ESR on large-area silicon MOSFETs. A newly discovered ESR signal at $g$ = 1.9999(1) is shown to arise from 2D electrons at the Si/SiO$_2$ interface; weakly confined electrons at gate voltages below threshold, and mobile 2D electron at gate voltages above threshold. Several experiments lead us to this assignment, including (1) the dependence of the signal on applied gate voltage, (2) an identical signal detected by EDMR above threshold, (3) the saturation of the signal intensity at gate voltages above threshold as expected for a 2DEG, and (4) a Curie-like temperature dependence of the signal below threshold as expected for confined electrons at the interface. As noted above, the temperature dependence of the spin density above threshold does not fit a simple Pauli law, and this behaviour is still under investigation.

This work was supported by the NSF through the Princeton MRSEC (DMR-0213706) and by NSA/LPS and ARO through LBNL (MOD713106A) and through the University of Wisconsin (W911NF-04-1-0389).

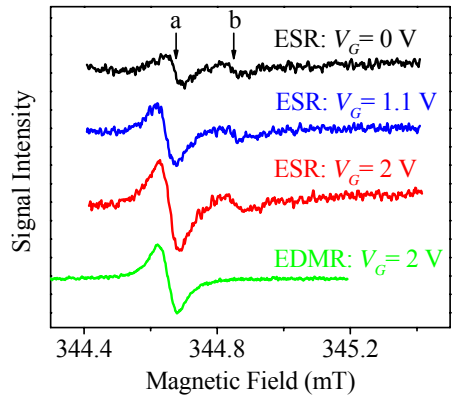

Fig. 1: ESR (top three) and EDMR (bottom) spectra for inversion MOSFET at 5 K with the magnetic field applied perpendicular to the 2D electron layer. Applied gate voltages ($V_G$) are as indicated for each spectrum. The strong signal at $g = 1.9999(1)$ labelled with the arrow 'a' is from 2D electrons. The weak signal at $g = 1.9988(1)$ labelled with the arrow 'b' is from conduction electrons in the source-drain contacts of the MOSFET.

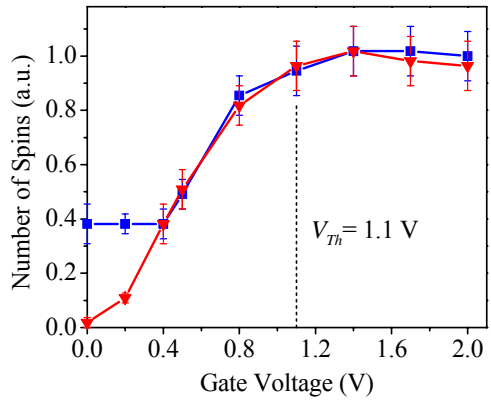

Fig. 2: Number of unpaired spins as a function of gate voltage for inversion MOSFET at 5 K. The squares (■) and triangles (▼) show the signal in the dark and after brief illumination with white light, respectively. The solid lines are a guide for the eye. The dashed line denotes the threshold voltage ($V_{Th}$ = 1.1 V).

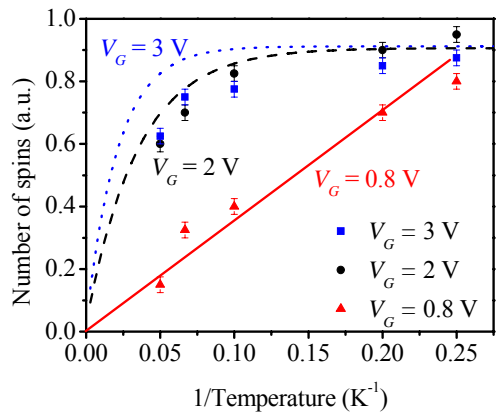

Fig. 3: Number of unpaired spins as a function of temperature for inversion MOSFET. The triangles (▲), circles (●) and squares (■) show the signal at gate voltages $V_G$ = 0.8, 2 and 3 V, respectively. Solid line is a linear fit to the $V_G$ = 0.8 V data showing a Curie law dependence. Dashed and dotted lines show Pauli law dependence for the $V_G$ = 2 and 3 V data, respectively.